%Paper: hep-th/9212068
%From: vpn@cuphyg.phys.columbia.edu (V.P. Nair)
%Date: Thu, 10 Dec 92 11:18:32 EST

\magnification=\magstep1
\baselineskip=18pt
\overfullrule=0pt
\nopagenumbers
\footline={\ifnum\pageno>1\hfil\folio\hfil\else\hfil\fi}
\font\twelvebf=cmbx12

\def\tr{{\rm tr}}
\def\Tr{{\rm Tr}}

\def\gA{{\gamma \cdot A}}
\def\gp{{\gamma \cdot p}}
\def\gq{{\gamma \cdot q}}
\def\gr{{\gamma \cdot r}}

\def\ap{{\alpha_p}}
\def\aq{{\alpha_q}}
\def\ar{{\alpha_r}}
\def\bp{{\beta_p}}
\def\bq{{\beta_q}}
\def\br{{\beta_r}}
\def\abp{{\bar \alpha_p}}
\def\abq{{\bar \alpha_q}}
\def\abr{{\bar \alpha_r}}
\def\bbp{{\bar \beta_p}}
\def\bbq{{\bar \beta_q}}
\def\bbr{{\bar \beta_r}}

\def\aoq{A_1 \cdot Q}
\def\atq{A_2 \cdot Q}
\def\athq{A_3 \cdot Q}
\def\kq{{k \cdot Q}}

\def\vqk{{\vec Q \cdot \vec k}}
\def\dq{{\cdot Q}}

\rightline{CU-TP-579}
\rightline{November 1992}
\vskip 1in
\centerline{\twelvebf Chern-Simons Theory and the Quark-Gluon Plasma}
\vskip .5in
\centerline{\it Ravit Efraty and V.P.Nair}
\vskip .1in
\centerline{Physics Department, Columbia University}
\centerline{New York, NY 10027}
\vskip .5in
\centerline{\bf Abstract}
\vskip .1in
\noindent  The generating functional for hard thermal loops in Quantum
Chromodynamics is important in setting up a resummed
thermal perturbation theory, so that all terms of a given order in the
coupling constant can be consistently taken into account. It is also the
functional which leads to a gauge invariant description of Debye
screening and plasma waves in the quark-gluon plasma. We have recently
shown that this functional is closely related to the eikonal for a
Chern-Simons gauge theory. In this paper, this relationship is explored
and explained in more detail along with some generalizations.

\vskip 1in
\noindent This research was supported in part by the U.S. Department of Energy.
\vfill\eject

\noindent
{\bf 1. Introduction}
\vskip .2in
We have recently shown that the generating functional for hard thermal
loops in Quantum Chromodynamics (QCD) is very closely related to the
Chern-Simons (CS) gauge theory $^1$. In this paper, we shall explore in
further detail this remarkable connection between QCD at finite
temperature and the CS theory.

Conceptually, the question we are considering is very simple. We
consider QCD at temperatures well into the deconfinement phase. Thus we
have a hot plasma of quarks and gluons. As is well known, one expects to
have Debye screening in a plasma. If we consider an Abelian plasma of
positive and negative charges $\pm e$, viz. electrodynamics,
screening can be understood using the classic argument of Debye.
One considers the Poisson equation for the electrostatic potential of a
test charge, say positive, in the plasma.
$$
\eqalignno{
-\nabla^2~A_0~&=~ ne\left( {{e^{eA_0 /T}~-~e^{-eA_0/T}}\over{e^{eA_0/T}~+~
e^{-eA_0/T}}}\right) &(1a)\cr
&\approx \left( ne^2\over T\right)&(1b) \cr}
$$
where the right hand side is the charge density in the vicinity of the
test charge. $n$ is the average number density of particles. The
exponentials are the Boltzmann factors giving the preferential
accumulation of negative and depletion of positive charges in the
vicinity due to the Coulomb forces, with proper normalization.
The approximation (1b), which is valid for high temperatures, shows that
the solutions have the screened Coulomb form ${1\over r}\exp(-m_D r)$ with
a Debye screening mass $m_D^2=(ne^2/T)$. For a relativistic plasma,
the qualitative features of this argument are valid and with $n\sim
T^3$, we expect $m_D^2\sim e^2 T^2$. And by calculating the photon
propagator in thermal electrodynamics, one can indeed obtain this
result $^2$.

The above argument is presented in terms of potentials and as such, it
does not seem to be gauge invariant. For the Abelian case, one can
easily reformulate the arguments using only the gauge invariant electric
and magnetic fields. However, in a non-Abelian plasma, such as the
quark-gluon plasma, it is difficult to avoid the use of gauge potentials
altogether. Further, since even the notion of the charge of a gluon has
to be defined with respect to some chosen Abelian direction of the gauge
group, it is clear that the simple argument of equation (1) will have to
be modified. The question of interest to us is: how do we obtain a gauge
invariant description of Debye screening in a non-Abelian plasma (in
terms of gauge potentials)?  More specifically, we need a functional of the
gauge potentials, $\Gamma [A]$, which is generated by the statistical
distributions and is effectively a gauge invariant mass term for the
gauge fields.
Screening, of course, is the static part of a more general problem, the
dynamical part of which is the propagation of plasma waves. Such a $\Gamma [A]$
will thus be
important for the discussion of plasma waves as well.

Hard thermal loops are closely related to the above. They arise because
of the need to carry out a partial resummation of perturbation theory in
thermal QCD. The need for resummation is easily seen using the following
argument due to Pisarski $^3$. Consider the elementary polarization diagram
of gluons, fig.(1a) and its first correction fig.(1b). Before the final
loop integration over $p$, the ratio of diagram (1b) to (1a) is $\Pi
(p)/p^2$. Because of Debye screening, the polarization tensor $\Pi (p)$
has a small $p$-expansion, $\Pi (p) \sim g^2 T^2 ~+~ g^2 T \vert
p\vert~+...$. ($g$ is the quark-gluon or gluon-gluon coupling constant.)
We thus see that for the small $p$-regime of integration, i.e. $p \leq
g^2T^2$, the naively higher order diagram (1b) is comparable to the
lower order term (1a). Therefore, to be consistent to a given order in
$g$, one must sum over (1a), (1b), and a series of further insertions of
$\Pi (p)$. This resummation leads to new effective propagators, and also
new effective vertices in general. The resummation can be summarized by
saying that the new propagators and vertices arise from an action
$$
S~=~ \int ~{\textstyle -{1\over 4}}F^2~+~\Gamma [A]\eqno(2)
$$
where we add a term $\Gamma [A]$ to the standard Yang-Mills action for
gluons.

Going back to figures (1), we see that the regime of interest is when
the momentum external to $\Pi (p)$, viz. $p$ is of the order $gT$ or
less. This holds in general and the momenta carried by the potentials in
$\Gamma [A]$ can be taken to be ${\buildrel < \over \sim} gT$.
General power counting
arguments by Pisarski and Braaten show that $\Gamma [A]$ is the sum
of `hard thermal loops' $^4$. These are one-loop diagrams where the
external momenta are ${\buildrel < \over \sim} gT$ and the loop
momentum is of the order of $T$, i.e. relatively hard.
An explicit form of $\Gamma [A]$, the generating functional of these
hard thermal loops, is necessary for the proper set up of thermal
perturbation theory. Although of seemingly different origin, $\Gamma
[A]$ so defined will be the same as the functional describing Debye
screening and plasma waves. We shall see that it is also closely related
to the Chern-Simons gauge theory.

The Chern-Simons (CS) action, it may be recalled, made its appearance in
physics literature over a decade ago as a mass term for gauge fields in
three dimensions $^5$. Studies since then have revealed a number of
fascinating features of this action. The Abelian version can be used for
spin transmutation, converting spin-zero bosons into anyons $^6$. The
correlators of Wilson lines in a pure CS theory are related to the
polynomial invariants of knot theory $^7$. Pure CS theory is also related to
conformal field theory and Wess-Zumino-Witten (WZW) models in two
dimensions $^{7,8}$.
Chern-Simons-Higgs theories have an intriguing set of vortex
solutions $^9$. Finally actions related to the CS action can be used for a
Lagrangian description of selfdual gauge fields and integrable systems
$^{10}$.
Our observation gives another context, viz. the very physical context of
the quark-gluon plasma, for the understanding  of which the CS theory is
very useful.

This paper is organized as follows. We begin with the explicit
evaluation of some hard thermal loops.
The two- and three-point functions are evaluated.
These calculations have some overlap
with the similar calculations of Frenkel and Taylor $^{11}$. However
we organize and orient the
calculations towards a more efficient realization of the CS connection.
In section 3, generalizations to $n$-point functions using the gauge
invariance of $\Gamma [A]$ are considered. The strategy of using
gauge invariance in this way is, to some extent, parallel to the work of
Taylor and Wong $^{12}$. In section 4, we discuss the pure CS theory
highlighting the eikonal and other features of relevance to the plasma
problem. In section 5, we obtain $\Gamma [A]$ in terms of the eikonal of
the CS theory. In fact, equation (86) is the central result of our
analysis.
Plasma waves are briefly discussed in section 6; non-Abelian plasmons,
we argue, must be understood as the propagating solutions of the
effective action (88).
\vskip .2in
\noindent
{\bf 2. Evaluation of Hard Thermal Loops}
\vskip .2in
We begin with the explicit evaluation of some of the hard thermal loops.
We shall consider one-loop quark graphs with two and three external gluon
lines; analysis of these diagrams will suffice to abstract many of the
features
that generalize to the $n$-point functions. As mentioned in the introduction,
there are many related but different formalisms for dealing with thermal
corrections in a field theory $^{13}$. We shall use Minkowski space propagators
with thermal averages for products of creation and annihilation operators.
The calculation is conceptually the simplest in this formalism. The
relevant part of the Lagrangian for the quark fields $q, \bar q$ is

$$ {\cal L} = \bar q ~i \gamma \cdot ( \partial + A) ~q  \eqno(3)$$

\noindent where $A_{\mu} = -i t^a A^a_{\mu}$ is the Lie algebra valued
gluon vector
potential, $t^a$ are hermitian matrices corresponding to the generators
of the Lie algebra in the representation to which the quarks belong.
We shall not explicitly display the quark-gluon coupling constant $g$,
as it is easily recovered at any stage by $A \rightarrow g A$.
The one-loop quark graphs are given by the effective action

$$ \Gamma = -i ~\Tr ~\log ~(1 + S ~\gamma \cdot A)  \eqno(4)$$

\noindent where

$$ S(x,y) = \langle T ~q(x) \bar q(y) \rangle  \eqno(5)$$

\noindent is the quark propagator. $\Tr$ in $(4)$ includes the functional trace
as usual. The two-gluon and three-gluon terms in $\Gamma$ are
given by

$$ \Gamma ^{(2)} = {i \over 2} \int d^4x d^4y~\Tr \left[ ~ \gA(x) S(x,y)
\gA(y) S(y,x) \right] \eqno(6a)
$$

$$ \Gamma ^{(3)} = -{i \over 3} \int d^4x d^4y d^4z ~\Tr \left[~
 \gA(x) S(x,y) \gA(y) S(y,z)
\gA(z) S(z,x) \right]  \eqno(6b)$$

\noindent where $\Tr$ denotes the trace over the color matrices ${t^a}$ and
the Dirac matrices.
The quark propagator is given by

$$\eqalign{ S(x,y) = \int {d^3p \over (2\pi )^3}{1 \over 2p^0} \lbrace
            &\Theta (x^0 - y^0) \lbrack \ap e^{-i p \cdot (x-y)} \gp
            +\bbp e^{i p' \cdot (x-y)}  \gp' \rbrack - \cr
            & \Theta (y^0- x^0) \lbrack \bp e^{-i p \cdot (x-y)} \gp
            +\abp e^{i p' \cdot (x-y)}  \gp' \rbrack \rbrace \cr}  (7) $$

\noindent where $p^0 = |\vec p|, \; p = (p^0,|\vec p|),
\; p' = (p^0,-|\vec p|),
$
and $\Theta(x)$, of course, is the step function. Also
$$ \ap = 1 - n_p,  ~~~~~~\bp = n_p.  \eqno(8)$$

The distribution functions $n_p, \bar n_p$ are defined by the thermal
averages
$$\eqalignno {\langle  a_p^{\dag \alpha,r} a_p^{\beta,s} \rangle& = n_p
   \delta^{rs} \delta^{\alpha \beta} \cr
   \langle  b_p^{\dag \alpha,r} b_p^{\beta,s} \rangle &=\bar n_p
   \delta^{rs} \delta^{\alpha \beta} &(9) \cr} $$

\noindent where $(a_p^{\dag \alpha,r} , a_p^{\alpha,r}),
        (b_p^{\dag \alpha,r} , b_p^{\alpha,r})$ are the annihilation and
creation operators for quarks and antiquarks respectively.
$\alpha,\beta$ are spin indices; $r,s$ are color indices.
For a plasma of zero fermion number, we can take
$$ n_p = \bar n_p = {1 \over e^{p^0/T} + 1}  \eqno(10) $$

For a plasma with a nonzero value of fermion number, there is a chemical
potential and correspondingly $n_p, \bar n_p$ are not equal.
In using the expression (7) for the propagator in (6), we encounter many terms
corresponding to different ordering of the time arguments $x^0, y^0, z^0.$
The strategy will be to carry out the time-integrations first,
introducing convergence
factors $e^{\pm \epsilon x^0},e^{\pm \epsilon y^0} $ etc.,$\epsilon$ small
and positive, as required. The integrations give energy denominators and
bring the result to a form where simplification appropriate to a hard thermal
loop, viz. the loop momentum being hard $(\sim T)$ and the external momenta
being relatively soft $(\sim g T),$ can be implemented easily.

\vskip .2in
\noindent
{\bf Simplification of the two-point function}
\vskip .2in
In using (7) for the quark propagator, we find four terms in $\Gamma^{(2)}$
with $x^0 > y^0 $ and four terms with $y^0 > x^0.$ Writing

$$ A_\mu (x) = \int {d^4k \over (2 \pi)^4} ~e^{ikx} ~A_\mu (k) \eqno(11)$$
and carrying out the time-integrations we get

$$ \eqalignno {\Gamma^{(2)} = -{\textstyle {1 \over 2}}
 \int d\mu(k)\int {d^3q \over (2\pi)^3} &{1\over 2 p^0}{1\over 2 q^0} \left[
{}~T(p,q) \bigl( {\ap \bq \over p^0-q^0-k^0-i\epsilon}-
{\aq \bp \over p^0-q^0-k^0+i\epsilon} \bigr)+\right. \cr
&\left. T(p,q') \bigl( {\ap \abq \over p^0+q^0-k^0-i\epsilon}-
{\bp \bbq \over p^0+q^0-k^0+i\epsilon} \bigr)+ \right. \cr
&\left. T(p',q) \bigl( {\abp \aq \over p^0+q^0+k^0-i\epsilon}-
{\bbp \bq \over p^0+q^0+k^0+i\epsilon} \bigr)+ \right. \cr
&\left. T(p',q') \bigl( {\abp \bbq \over p^0-q^0+k^0-i\epsilon}-
{\bbp \abq \over p^0-q^0+k^0+i\epsilon} \bigr) \right]  &(12) \cr} $$

\noindent where

$$T(p,q) = \Tr \lbrack ~\gA(k) \gp \gA(k') \gq ~\rbrack \eqno(13)$$

\noindent and

$$d\mu (k) = (2\pi)^4 \delta^{(4)} (k+k'){d^4k \over (2\pi)^4}
             {d^4k' \over (2\pi)^4} \eqno(14)$$

In (12), $\vec p = \vec q + \vec k.$ Since $p^0 = |\vec q + \vec k |
\simeq q^0  + \vec q \cdot \vec k/q^0$ for $|\vec k|$ small
compared to $|\vec q|,$ the denominators in (12) involve $ k \cdot Q,
\; k \cdot Q'$ and $2 q^0 + k \cdot Q, \; 2 q^0 + k \cdot Q'$ where

$$ Q= (1, ~{\vec q \over q^0}), \; Q' = (1, ~-{\vec q \over q^0})
\eqno (15)$$

The $i \epsilon$'s in the denominators in (12)
can be let go to zero at this stage. Of course, this requires that the
kinematics be so chosen that $k\cdot Q,~k\cdot Q'$ and $q^0$ are not
zero. Making this choice and using $\alpha,\beta$
from (8), we find, for the temperature-dependent part of $\Gamma^{(2)},$

$$\eqalignno{\Gamma^{(2)} = -{\textstyle{1 \over 2}} \int
d\mu(k)\int {d^3q \over (2\pi)^3}&{1\over 2 p^0}{1 \over 2 q^0} ~\left[
(n_q-n_p){T(p,q) \over p^0-q^0-k^0}
+ (\bar n_q-\bar n_p){T(p',q') \over p^0-q^0+k^0}-\right. \cr
&\left. (n_p+\bar n_q){T(p,q') \over p^0+q^0-k^0}
-(\bar n_p+n_q){T(p',q) \over p^0+q^0+k^0} \right]. &(16) \cr}$$

We see that the result is linear in the distribution functions, a property
that is expected to be true in general $^{4}$.
Equation (16) agrees with Frenkel and Taylor $^{11}$. Some of the further
simplification of (16) can be made along the lines of their calculation.
For $|\vec k|$ small compared to the loop momentum $|\vec q|,$
we have

$$ p^0 -q^0 -k^0 \simeq - \kq ~~~~~~~~ p^0 -q^0 +k^0 \simeq  \kq'$$
$$ p^0 +q^0 \pm k^0 \simeq 2 q^0  \eqno (17)$$

$$T(p,q) \simeq 8 q^0 \tr (\aoq \atq)$$
$$T(p',q') \simeq 8 q^0 \tr (\aoq' \atq')   \eqno(18)$$
$$T(p',q) \simeq T(p,q') \simeq 4 q^0 \tr (\aoq' \atq + \aoq \atq' -
2 A_1 \cdot A_2)$$

\noindent where $A_1 = A(k), \; A_2 = A(k') $ and the remaining trace
in the expressions
for $T$'s, denoted by $\tr,$ is over color indices. The difference of
distribution functions can also be approximated as $n_p - n_q \simeq {dn \over
dq^0} \vqk.$ Using these results (16) simplifies to

$$\eqalignno{ \Gamma^{(2)} = -{\textstyle {1 \over 2}}
 \int d\mu(k)\int {d^3q \over (2\pi)^3}
{}~&\tr \Biggl[~ \bigl( {dn \over dq^0} {\aoq \atq \over \kq}-
{d\bar n \over dq^0} {\aoq' \atq' \over \kq'}\bigr) 2\vqk \cr
&\left. -{n +\bar n \over q^0}(\aoq' \atq + \aoq \atq' -
2 A_1 \cdot A_2) \right]  & (19) \cr } $$

\noindent
We have the result

$$\int d^3q {dn \over dq^0} f(Q) = -\int d^3q {2n \over q^0} f(Q)
  \eqno(20)$$
for any function $f$ of $Q,$ or $Q'.$ We can further use $ 2\vqk = \kq' -\kq.$
Expression (19) then simplifies to

$$\eqalignno{\Gamma^{(2)} = -{\textstyle {1 \over 2}} \int d\mu(k)\int
{d^3q \over (2\pi)^3}~\tr \Biggl[~
& {n +\bar n \over q^0}(2\aoq \atq - \aoq \atq' - \aoq' \atq +2 A_1\cdot
 A_2) \cr
&\left. -{n \over q^0} 2\aoq \atq {\kq' \over \kq} -
{\bar n \over q^0} 2\aoq' \atq' {\kq \over \kq'} \right]. &(21)\cr}$$

The angular integration in (21) over the directions of $\vec q$ (or $\vec Q$)
help simplify it further by virtue of

$$ \int d \Omega ~(2 \aoq \atq - \aoq \atq' - \aoq' \atq +2 A_1 A_2)
   =  \int d\Omega~ (2 \aoq \atq')  \eqno(22)$$
Defining

$$A_+ = {A\dq \over 2},~~~~~~~~ A_- = {A\dq' \over 2}  \eqno(23) $$

\noindent we can write (21) as

$$\Gamma^{(2)} = -{\textstyle {1 \over 2}} \int d\mu(k)
\int {d^3q \over (2\pi)^3}
{1 \over 2 q^0} ~16 ~\tr \left[~ A_{1+}A_{2-} (n+ \bar n) -
n {\kq' \over \kq} A_{1+}A_{2+}-
\bar n {\kq \over \kq'} A_{1-}A_{2-} \right].  \eqno(24) $$

This expression becomes more transparent when written in
coordinate space and finally using a Wick rotation to Euclidean
space. We define the Green's functions

$$G(x_1, x_2)=\int {e^{-ip \cdot (x_1-x_2)} \over p \cdot Q}
{d^4p \over (2\pi)^4}$$
$$G'(x_1, x_2)=\int {e^{-ip \cdot (x_1-x_2)} \over p \cdot Q'}
{d^4p \over (2\pi)^4}    \eqno(25)$$
In terms of the null vectors $Q = (1, \vec q/q^0), \;
Q' = (1,-\vec q/q^0), $ we can introduce the lightcone coordinates
$(u,v,x^T)$ as

$$u = {Q' \cdot x \over 2} , ~~~v = {Q \cdot x \over 2},~~~
\vec Q \cdot \vec x^T = 0  \eqno(26) $$

\noindent where $\vec Q = \vec q/q^0. $ We then have $Q \cdot \partial =
\partial_u,
\; Q' \cdot \partial = \partial_v.$ We shall also introduce a Wick rotation
to Euclidean coordinates by the correspondence

$$ 2v \leftrightarrow z, ~~~~~2u \leftrightarrow \bar z $$
$$\partial_u =  Q \cdot \partial \leftrightarrow 2 \partial_z,~~~~
\partial_v =  Q' \cdot \partial \leftrightarrow 2 \partial_{\bar z} \eqno(27)$$
The Green's functions in (25) are the continuations of the Euclidean
functions

$$G_E(x_1,x_2) = {1 \over 2\pi i} {\delta^{(2)}(x_1^T-x_2^T)  \over
(\bar z_1-\bar z_2)}$$
$$G'_E(x_1,x_2) = {1 \over 2\pi i} {\delta^{(2)}(x_1^T-x_2^T)  \over
( z_1-z_2)}    \eqno (28)$$
This leads to the correspondence

$$ {\kq' \over \kq} \longleftrightarrow {1 \over \pi}~{1 \over \bar z_{12} \bar
z_{21}}$$
$$ {\kq \over \kq'} \longleftrightarrow {1 \over \pi}~{1 \over z_{12} z_{21}}$$
$$ {1\over \kq} \longleftrightarrow {1 \over  2 \pi i}~{1 \over \bar z_{12}}
 \eqno(29)$$

\noindent where $z_{ij} = z_i -z_j,$ etc. Equation (24) for $\Gamma^{(2)}$ can
then be written as

$$\eqalignno{\Gamma^{(2)} &= -{\textstyle {1 \over 2}}
\int {d^3q \over (2\pi)^3}
{1 \over 2q^0} ~16
{}~\tr \left[ ~(n + \bar n)~\int d^4x~A_+(x)A_-(x) \right.\cr
&\left. -n ~\pi \int {d^2x^T}{d^2z_1\over \pi} {d^2z_2\over \pi}
{A_+(x_1)A_+(x_2) \over \bar z_{12}\bar z_{21}}-
\bar n ~\pi \int {d^2x^T}{d^2z_1\over \pi}{d^2z_2\over \pi}
{A_-(x_1)A_-(x_2) \over z_{12}z_{21}}
\right]. & (30)\cr} $$

We define the functional $I(A_+)$ by the formula
$$\eqalignno{ I(A_+) &= i \sum {(-1)^n \over n}\int ~
d^2x^T~{d^2z_1\over \pi}\cdots {d^2z_n \over \pi}~{\tr(A_+(1) \cdots A_+(n))
\over \bar z_{12}  \bar z_{23} \cdots \bar z_{n1} }
\cr
&= {i\over 2} \int ~{d^2x^T}~{d^2z_1 \over \pi}
{d^2z_2 \over \pi} ~{\tr(A_+(1)A_+(2)) \over
\bar z_{12} \bar z_{21} }
 + \cdots &(31) \cr}$$
We shall show later that $I(A_+)$ is related to the eikonal for a
Chern-Simons theory. Equation (30) for $\Gamma^{(2)}$ can finally
be written as
$$\Gamma^{(2)} = \int {d^3q \over (2\pi)^3}{1 \over 2q^0}
                   ~~K^{(2)}[A_+,A_-]     \eqno(32a)$$

\noindent where
$$ K[A_+,A_-] =-16 ~\left[~ {(n +\bar n)\over 2} \int d^4x~\tr ~\bigl(
A_+(x)A_-(x)\bigr)
+n ~i \pi  I(A_+) + \bar n ~i \pi \tilde I(A_-) \right]   \eqno(32b)$$

\noindent $K^{(2)}[A_+,A_-]$ in (32a) denotes terms in $K$ which are quadratic
in $A.$
$\tilde I $ is obtained from $I$ by $z \leftrightarrow \bar z.$

Although we have introduced different distribution functions $n, \bar n$
for quarks and antiquarks, it is only $n + \bar n$ which is relevant at
high temperatures. We see that, by virtue of $\int d\Omega ~I(A_+) =
\int d\Omega ~\tilde I(A_-),$ we can write (32b) as

$$ K[A_+,A_-] =-16 ~{n + \bar n \over 2}~\left[~ \int d^4x
{}~\tr \bigl( A_+(x)A_-(x) \bigr)
+i \pi I(A_+) + i \pi \tilde I(A_-)  \right] . \eqno(33)$$

The integral over the magnitude of $\vec q$ in (32) can be easily
carried out. With zero chemical potential,
$$\Gamma^{(2)} = {-T^2 \over 12\pi} \int d\Omega ~\left[~ \int
d^4x~\tr \bigl( A_+ ~A_- \bigr)
+i \pi I^{(2)}(A_+) + i \pi \tilde I^{(2)}(A_-)  \right]. \eqno(34)$$
\vskip .2in
\noindent
{\bf Simplification of the three-point function}
\vskip .2in
When expression (7) for the propagator is used in (6b) for the
three-point function and the time-integrations are carried out we get
a number of terms with different types of energy-denominators. The
dominant terms will have differences of energies $p^0, q^0, r^0$
corresponding to the three propagators so that for a small $k,$
we get soft denominators. The most important terms will have
differences of momenta in both the energy denominators. There are six
orderings of the time labels $x^0, y^0, z^0$ and for each ordering we
get two terms with both denominators soft; one of these terms will involve
quark distribution factors $\alpha, \beta$ and the other involves
the antiquark distribution factors $\bar \alpha, \bar \beta.$
By looking at the exponentials from the propagators with maximal
differences of momenta, we can easily write down these terms. As
in the case of the two-point function we shall neglect the
$i \epsilon$-factors.
The result is then

$$ \Gamma^{(3)} = {-i \over 3} \int d\mu(k)~~\Gamma (k_1,k_2,k_3)   \eqno(35)$$

$$\eqalignno{ \Gamma (k_1,k_2,k_3) =& \int {d^3q \over (2\pi)^3} {1\over 2 p^0}
{1\over 2 q^0} {1\over 2 r^0} \bigl\{ T(p,q,r) \lbrack
{\ap \aq \br- \bp \bq \ar \over (p^0-r^0-k_1^0)(q^0-r^0+k_3^0)}+\cr
&{\bp \aq \br- \ap \bq \ar \over (p^0-r^0+k_2^0)(q^0-r^0+k_3^0)}+
{\bp \aq \ar- \ap \bq \br \over (p^0-q^0+k_2^0)(p^0-r^0-k_1^0)} \rbrack +\cr
&T(p',q',r') \lbrack
{\bbp \bbq \abr- \abp \abq \bbr \over (p^0-r^0+k_1^0)(q^0-r^0-k_3^0)}+\cr
&{\abp \bbq \abr- \bbp \abq \bbr \over (p^0-q^0-k_2^0)(q^0-r^0-k_3^0)}+
{\abp \bbq \bbr- \bbp \abq \abr \over (p^0-q^0-k_2^0)(p^0-r^0+k_1^0)}
\rbrack  \bigr\}.  & (36)\cr}$$
Using the expressions for $\alpha, \beta, \bar \alpha, \bar \beta$ in
terms of $n, \bar n$ we can further simplify these terms as
$$\eqalignno{\Gamma (k_1,k_2,k_3) &= \int {d^3q \over (2\pi)^3} {1\over 2 p^0}
{1\over 2 q^0} {1\over 2 r^0} \bigl\{
T(p,q,r) \lbrack
{n_p \over (p^0-r^0-k_1^0) (p^0-q^0+k_2^0)}+\cr
&{n_q \over (p^0-q^0+k_2^0)(r^0-q^0-k_3^0)}+
{n_r \over (p^0-r^0-k_1^0)(q^0-r^0+k_3^0)}\rbrack \cr
&-T(p',q',r') \lbrack
{\bar n_p \over (p^0-r^0+k_1^0) (p^0-q^0-k_2^0)}+\cr
&{\bar n_q \over (p^0-q^0-k_2^0)(r^0-q^0+k_3^0)}+
{\bar n_r \over (p^0-r^0+k_1^0)(q^0-r^0-k_3^0)}
\rbrack  \bigr\} .  & (37)\cr}$$
In these equations, $\vec p = \vec q - \vec k_2$ and
$\vec r = \vec q + \vec k_3$
and

$$d\mu (k) = (2\pi)^4 \delta^{(4)} (k_1+k_2+k_3)~{d^4k_1 \over (2\pi)^4}
             {d^4k_2 \over (2\pi)^4}{d^4k_3 \over (2\pi)^4} \eqno(38)$$

\noindent The numerators in (36,37) are given by
$$T(p,q,r) = \Tr ~\left[~ \gA_1 \gp \gA_2 \gq \gA_3 \gr \right]
            \simeq 16 {q^0}^3 ~\tr \bigl(\aoq \atq \athq \bigr) \eqno(39)$$

\noindent where $A_1 = A(k_1),$ etc.

Using the approximation as in (17) for the energy-denominators, we get

$$\eqalignno{\Gamma (k_1,k_2,k_3)& = \int {d^3q \over (2\pi)^3} \bigl\{
-2 \tr (\aoq \atq \athq) \lbrack {n_p \over k_1 \cdot Q k_2
\cdot Q} + {n_q \over
 k_2 \cdot Q k_3 \cdot Q} + {n_r \over k_3\cdot Q k_1 \cdot Q} \rbrack  \cr
&+~2~ \tr (\aoq' \atq' \athq') \lbrack
{\bar n_p \over k_1 \cdot Q k_2 \cdot Q} +
{\bar n_q \over k_2 \cdot Q k_3 \cdot Q} + {\bar n_r \over k_3\cdot Q k_1 \cdot
 Q} \rbrack \bigr\}.  &(40)\cr}$$
Further, we can approximate the distribution functions $n_p, n_r$ as

$$n_p \simeq n_q + {dn \over dq^0}(-\vec k_2 \cdot \vec Q)
\simeq n_q + {1 \over 2} {dn \over dq^0}(k_2 \cdot Q -k_2 \cdot Q') $$
$$n_r \simeq n_q + {dn \over dq^0}(\vec k_3 \cdot \vec Q)
\simeq n_q + {1 \over 2} {dn \over dq^0}(k_3 \cdot Q' -k_3 \cdot Q)
\eqno(41) $$
When these expressions are used in (40), terms with no ${dn \over  dq^0}\;$
factor is zero by momentum conservation. With one partial integration
over $q^0 = |\vec q|,$ the remaining terms simplify as

$$\eqalignno{\Gamma (k_1,k_2,k_3) = -32 \int {d^3q \over (2\pi)^3}
&{1 \over 2q^0} \bigl\{
n~ \tr (A_{1+}A_{2+}A_{3+}) [ {k_2 \cdot Q' \over k_1 \cdot Q k_2
\cdot Q}
 - {k_3 \cdot Q' \over k_1 \cdot Q k_3 \cdot Q}] +\cr
&\bar n ~\tr (A_{1-}A_{2-}A_{3-}) \lbrack {k_2 \cdot Q \over k_1 \cdot Q'
k_2 \cdot Q'}  - {k_3 \cdot Q \over k_1 \cdot Q' k_3 \cdot Q'} \rbrack \bigr\}.
  &(42) \cr}$$
By using the correspondence rules (29) we can write (42) and $\Gamma^{(3)}$
of (35) as

$$\eqalignno{\Gamma^{(3)} &= {-16\pi \over 3} \int
{d^3q \over (2\pi)^3}{1 \over 2q^0}~d^2x^T~{d^2z_1 \over \pi}{d^2z_2
\over \pi} {d^2z_3 \over \pi}
\left[~
n {\tr(A_+(x_1)A_+(x_2)A_+(x_3)) \over \bar z_{12}\bar z_{23}\bar z_{31}}+
\right. \cr
&\left. ~~~~~~~~~~~~~~~~~~~~~
\bar n {\tr(A_-(x_1)A_-(x_2)A_-(x_3)) \over z_{12} z_{23} z_{31}}
\right] \cr
&=\int {d^3q \over (2\pi)^3}{1 \over 2q^0}~ (-16) \left[~
n~i \pi I^{(3)}(A_+) + \bar n ~i \pi \tilde I^{(3)}(A_-) \right], &(43)\cr}$$

\noindent where $I^{(3)}$ denotes the terms of cubic order in the $A$'s in the
expansion (31) of $I(A).$ Again by virtue of $\int d\Omega ~I(A_+) =
\int d\Omega ~I(A_-),$ the result depends only on $n +\bar n.$ We can
then combine equations (32) for the two-point function and (43)
for the three-point function as

$$\Gamma = \int {d^3q \over (2\pi)^3}{1 \over 2q^0} ~~K[A_+, A_-]  \eqno(44)$$

\noindent where $K$ is given by equation (33) and we retain terms
up to the cubic order in $A$'s.

We shall now rewrite this result in a slightly different way, with a view to
generalizations. Using the identity

$$\int d\Omega ~\tr(A_+A_-) = -{\textstyle{1 \over 2}}~\left[~2\pi A_0^a A_0^a
-
\int d\Omega ~(A_+^aA_+^a) \right]  \eqno(45) $$

\noindent and  $\int d\Omega ~\tilde I(A_-) = \int d\Omega ~I(A_+),$
we can write
$$\Gamma =  {1\over (2\pi)^3}\int q dq~ {(n +\bar n) \over 2}
\left[ \int d^4x~2\pi A_0^a A_0^a  - \int d\Omega~d^4x~ (A_+^aA_+^a)
- 4 \pi i I(A_+) \right].  \eqno(46) $$

\noindent Notice that $\Gamma,$ except for
the $A_0^a A_0^a$ term, depends on $A_\mu$
only through the combination $A_+ = A \cdot Q/2.$ (Here $q=|\vec q |$.)

It is possible to interpret the various terms in equation (37) in
terms of certain forward scattering amplitudes. If we use an interaction
Hamiltonian $H_{int}= \gamma \cdot A \gamma \cdot P$ and calculate the
forward scattering amplitude, using standard time-dependent perturbation
theory, for a quark of momentum $p$, we get the
first term in (37). The sum of this amplitude over spin and colors gives
the traces. Integration over all momenta, distributed according to
$n_p$, gives the  thermal loop contribution. The other terms in (37) can
be similarly interpreted in terms of permutations of the external
gluons. (This interpretation generalizes to four-point and higher
functions as well.) $^{14}$
\vskip .2in
\noindent
{\bf 3. Generalizations}
\vskip .2in
We shall now consider some of the general features of the calculations
presented so far. In generalizing to four-point and higher functions,
it is easy to see which kind of terms will dominate. We need the maximal
number of soft denominators. This will come from the time-orderings which give
maximal number of differences of momenta in the exponential for the
propagators.
In an $n$-point function, we can get $(n-1)$ such terms.
The denominators simplify
to products of $\kq$'s (and $\kq'$'s for the antiquark distributions). The
general result has the structure

$$ \Gamma = {i^n \over n}  \int {d^3p_1 \over (2\pi)^3} \prod_i  {1 \over
2p_i^0}
{}~\left[ {T(p_1, p_2, \cdots,p_n) n_{p_1} \over
k_2 \cdot Q \cdots k_{(n)} \cdot Q}
+{\rm cyclic} \right]d\mu (k_1, \cdots,k_n) \eqno(47)$$

\noindent and a similar term for the antiquark distributions. (We have renamed
$k$'s compared to (40).) The numerator, in the kinematic regime of
interest, viz. $k$'s small compared to $p,$ becomes $2^{n+1}p_0^n \aoq
\atq \cdots A_n \cdot Q$. We can thus expect the final result to depend on
$A$'s only through $A \cdot Q$ for quarks
(and $A \cdot Q'$ for the antiquarks).
Since $\int d\Omega ~f(A \cdot Q)  = \int d\Omega ~f(A \cdot Q'),$ in fact
we can take all higher terms to be a function of $A \cdot Q.$
Further, since
$$ \sum_{cyclic} {1 \over k_2\dq k_3\dq  \cdots k_{n} \dq} = 0 \eqno(48)$$

\noindent by conservation of energy and momentum, the nonzero contributions
involve ${dn \over dp^0}$ and hence the quantity is proportional to
$\int {d^3p \over (2\pi)^3} {n_p \over 2p^0}$ or $T^2.$ We may
then write the result for higher order terms in the form

$$\Gamma_{higher} = \int {d^3q \over (2\pi)^3}{1\over {2q^0}}
{}~{(n + \bar n) \over 2} ~W(A\dq). \eqno(49)
$$
Combining this with (46), we see that we can write $\Gamma$ as

$$\Gamma =  {1\over (2\pi)^3}\int q dq~ {(n +\bar n) \over 2}
\left[ \int d^4x~ 2\pi A_0^a A_0^a +\int d\Omega ~W(A \dq)
 \right].  \eqno(50) $$
The key observation here is that, including all higher powers of $A_{\mu},$
$W$ depends on $A_{\mu}$ only through $A\cdot Q.$

Consider now the contribution of gluon loops. The diagram (4) will generate
an $A_0^a A_0^a$-term, as in the case of quark loops. For the higher functions,
the dominant terms will come from the three-gluon vertices. The basic vertex
has the form
$$\eqalignno{V_{\alpha, \beta} &= 2 i A \cdot p \delta_{\alpha \beta} +
(2 i A \cdot k \delta_{\alpha \beta} + 4 i k_\alpha A_\beta) \cr
&\simeq 2i A \cdot p  \delta_{\alpha \beta} &(51)\cr}$$

\noindent where $p$ will be taken as the loop momenta and $A$ denotes
the external gluon of momentum $k.$ The color structure is
given by $A = -i T^a_{bc}A^a = -f^a_{bc}A^a,$ where $f^a_{bc}$ are
the structure constants of the gauge group. We then see that
gluon loops contribute the same way as quark loops except that the
color trace is over the $T^a$'s which are matrices in the adjoint
representation. Now, from the structure of the higher terms in
$A \cdot Q$ we see that, although we have factors like
$\tr (T^{a_1} T^{a_2}  \cdots T^{a_n}),$ all the matrix
products can be reduced by the symmetry properties of the
spacetime part to commutation rules and hence products of
structure constants except for one last trace over two $T$'s. Thus
the ratio of the gluon loops to quark loops is given by $C_A/C_F$
where for any representation $R$, $\tr~(t^at^b)_R=C_R\delta^{ab}$.
For gauge group $SU(N)$ with quarks in the fundamental representation, we
get $C_A/C_F=2N$. The fact that there are only two physical
polarizations of the gluon is taken care of by
assigning distribution functions only to the transverse physical
polarizations. Equivalently we may subtract a ghost-loop
contribution. (Naively, (51) leads to $2^n~\tr~\delta_{\mu\nu}=
2^n\cdot 4$ for an $n$-th order term. Actually, since there are only two
physical polarizations we get $2^{n+1}$ as for fermions.)
Further there are no separate $n$ and ${\bar n}$ contributions for
gluons. Putting all this together we get for the gluon contribution
$$
\Gamma=2N{1\over2\pi}\int {n_g\over2} q dq ~\left[~\int 2\pi
A_0^aA_0^a d^4+\int d\Omega ~W(A\cdot Q)\right]\eqno(52)
$$
\noindent where $n_g$ is the gluon distribution
$$n_g={1\over e^{q/T}-1}.\eqno(53)
$$
Strictly speaking we have not proved that
$W(A\cdot Q)$ for gluon loops in (52) is the
same as
$W(A\cdot Q)$ of the quark loops in (50).
The vertex (51) only shows that
the higher order contributions, for gluon loops also, depend only on
$A\cdot Q$. The strategy will be to
determine $W(A\cdot Q)$ by gauge invariance of the
$\Gamma$. $W$ is thus determined by the gauge transformation
properties of $A_0^aA_0^a$ and therefore will be the same for quark
loops and for gluon loops.
With zero chemical potential for the quarks we have
$$
\eqalignno{\int qdq~{(n+{\bar n})\over 2}&={\pi^2T^2\over12}\cr
\int q dq~ n_g  &={\pi^2T^2\over 6}&(54)\cr}
$$
\noindent We can combine (52) and (50) for $N_F$ flavors of
quarks as
$$\Gamma=
(N+ {\textstyle {1\over 2}}N_F)~{T^2\over 12\pi}~\left[\int d^4x~
2\pi A_0^aA_0^a
+\int d\Omega ~W(A\cdot Q)\right]. \eqno(55)
$$

Our arguments so far have been confined to one-loop
diagrams. The crucial result which can be proved by a
power-counting analysis of diagrams is that only one-loop
diagrams give hard thermal contributions. Thus in constructing
$\Gamma$ only one-loop diagrams need be considered $^4$. The other crucial
property of $\Gamma$ which helps us to proceed further is that $\Gamma$,
considered as a functional of $A_\mu$, is invariant under
gauge transformations of $A_\mu$ and further that $\Gamma$ is
independent of the gauge-fixing used for gluon propagators
in the internal lines. This property can be established by an
analysis of the Ward identities. A set of identities for the
gauge dependence of the generating functional for one-particle
irreducible or proper vertices can be derived using the BRST symmetry.
The thermal power counting rules, applied to the diagrammatic expansion
of various terms in these identities, show that the gauge dependence can
affect only terms which are of the order of $T$ or less, in a high
$T$-expansion. The leading term which is of order $T^2$, viz. the
generator of hard thermal loops, is gauge invariant and independent of
the gauge-fixing chosen for the internal lines $^{15}$.
We shall not repeat this analysis
here but notice that since only one-loop diagrams are important for
$\Gamma$, the thermal contribution is really classical and hence
the BRST Ward identities can be expected to reduce to a statement
of gauge invariance.

We know that, apart from the $A_0^aA_0^a$-term, $\Gamma$ can be taken to
depend on $A\cdot Q$ with a final integration over the orientations of
$\vec Q$. This result, viz.~(55), holds to
all orders in $A_\mu$. Of course, explicitly, we have calculated
$W(A\cdot Q)$ to cubic order in $A_\mu$. We can now use gauge
invariance along the lines of reference 12, to determine $W(A\cdot Q)$.
The condition for gauge invariance of $\Gamma$ is
$$
\int d\Omega~\delta W=4\pi\int d^4x ~{\dot A}_0^a \omega^a\eqno(56)
$$
\noindent where ${\dot A}_0$ is the time-derivative of $A_0^a$
and $\omega=-it^a\omega^a$ is the parameter of the gauge
transformation, i.e. $\delta A_\mu=\partial_\mu \omega
+[A_\mu,\omega]$. Equation
(56) can be realized by
$$
\delta W=\int d^4x ~A^a\cdot Q \omega^a.\eqno(57)
$$
\noindent One can check that (57) is indeed the way gauge invariance
is realized by analysis of the diagrams. It is clearly so for
the two- and three- point functions from our explicit calculations.
We now rewrite (57), using
$$\delta W=-\int d^4x(Q\cdot\partial{\delta W\over \delta(A\cdot Q)}
+[A\cdot Q,{\delta W\over\delta(A\cdot Q)}])^a\omega^a\eqno(58)
$$
\noindent as
$$
{\partial f\over\partial u}+[A\cdot Q,f]+{1\over 2}
{\partial(A\cdot Q)\over \partial v}=0,\eqno(59)
$$
\noindent where
$$
f={\delta W\over\delta(A\cdot Q)}+{\textstyle{1\over 2}}A\cdot Q.\eqno(60)
$$

\noindent We have used the lightcone coordinates from equation (26).
We now introduce the fields
$$
a_z={\textstyle{1\over 2}}A\cdot Q=A_{+}~~~~~a_{\bar z}=-f
=-{1\over2}{\delta W\over\delta a_z}-a_z\eqno(61)
$$
\noindent and also make the Wick rotation, as in (27).
The condition of gauge invariance, equation (59),
then becomes
$$
\partial_{\bar z}a_z-\partial_z a_{\bar z}+[a_{\bar z},a_z]=0.
\eqno(62)
$$

If $a_z$, $a_{\bar z}$ are thought of as the gauge
potentials of another gauge theory, we see that
equation (62) is the vanishing of the field strength or
curvature $F_{z{\bar z}}$. The gauge theory whose
equations of motion say that the field strengths vanish is
the Chern-Simons theory.
We shall therefore turn to a digression on the
Chern-Simons theory, returning to (62) and its solution in
section 5. Of course, an understanding of
Chern-Simons theory is not
absolutely essential to solving (62). (For a solution in a nonthermal
context, see ref.16.) One can simply
solve (62) and regard Chern-Simons theory as an interpretation of
the mathematical steps along the way. However Chern-Simons
theory does illuminate many of the nice geometrical properties
of the final result and is a worthwhile
digression.
\vskip .2in
\noindent
{\bf 4. Chern-Simons Theory}
\vskip .2in
The Chern-Simons theory is a gauge theory in two
space (and one time) dimensions. The action is given by
$$
S={k\over 4\pi}\int_{M\times[t_i,t_f]}d^3x~\epsilon^{\mu\nu\alpha}
{}~\tr(a_\mu\partial_\nu
a_\alpha+{\textstyle{2\over3}}a_\mu a_\nu a_\alpha).
\eqno(63)
$$
\noindent Here $a_\mu$ is the Lie algebra valued gauge potential,
$a_\mu=-it^a a_\mu^a$.
We shall consider $SU(N)$ gauge group in what follows.
$k$ is a constant whose precise value we do not need to specify
at this stage. We shall consider the spatial manifold to
be ${\bf R}^2$, or ${\bf C}$ since we shall be using complex
coordinates $z=x+iy$, ${\bar z}=x-iy$.
(Actually, we have sufficient regularity conditions at
spatial infinity that we may take $M$ to
be the Riemann sphere.) The equations of motion for the theory are
$$
F_{\mu\nu}=0.\eqno(64)
$$

The theory is best analyzed, for our purposes, in the
gauge where $a_0$ is set to zero. In this gauge, the
equations of motion (64) tell us that $a_z$, $a_{\bar z}$
are independent of time, but must satisfy the constraint
$$
F_{{\bar z}z}\equiv\partial_{\bar z}a_z-
\partial_z a_{\bar z}+[a_{\bar z},a_z]=0.\eqno(65)
$$
\noindent This constraint is just the Gauss law of the CS
gauge theory. It can be solved for $a_{\bar z}$ as a function
of $a_z$, at least as a power series in $a_z$. The result
is
$$
a_{\bar z}=\sum(-1)^{n-1}
\int {d^2z_1\over\pi}\cdots{d^2z_n\over\pi}~
{a_z(z_1,{\bar z}_1)a_z(z_2,{\bar z}_2)\ldots
a_z(z_n,{\bar z}_n)\over
({\bar z}-{\bar z}_1)({\bar z}_1-{\bar z}_2)\ldots({\bar z}_n-{\bar z})}.
\eqno(66)
$$
\noindent This can be easily checked using $\partial_z ({1\over {\bar z}-
{\bar z}'})=\pi\delta^{(2)}(z-z')$.

In the $a_0=0$ gauge, the action becomes
$$
S={ik\over\pi}\int dt d^2x ~\tr (a_{\bar z}\partial_0 a_z).\eqno(67)
$$
\noindent This shows that $a_{\bar z}$ is essentially canonically
conjugate to $a_z$. In fact in carrying out a variation
of $S$, we find the surface term $\theta(t_f)-\theta(t_i)$, where
$$
\theta={ik\over\pi}\int_Md^2x~\tr(a_{\bar z}\delta a_z).\eqno(68)
$$
\noindent (We assume $a_{\bar z}\delta a_z$ to vanish at spatial infinity.)
$\theta$ is the canonical one-form of the CS theory. (This is so by
definition; the canonical one-form in any theory can be defined by
$\theta$, where the surface term in the variation of the action is
$\theta_f ~-~\theta_i$, the subscripts referring to the final and
initial data surfaces. This is an old result going back to the theory of
canonical transformations and Hamilton-Jacobi theory. In more
modern times, for example, Schwinger's action
principle is essentially based on this statement. For some recent
references, see ref.17. )
$\theta$ is the analogue of $p_idx^i$ of point-particle mechanics; for an
action $S=\int dt dx~[{m{\dot x}^2\over2}-V(x)],$ we would find
$\theta=m{\dot x}_i\delta x^i=p_i\delta x^i$. We can make
another variation of $\theta$, antisymmetrized with respect to the
variation $\delta a_z$, denoted by the wedge product sign,
and write
$$\eqalignno{\omega \equiv \delta\theta&={ik\over\pi}
\int_Md^2x ~\tr(\delta a_{\bar z}\wedge\delta a_z)\cr
&={\textstyle{1\over2}}\int d^2x d^2x'
{}~\omega_{ab}(x,x')~\delta\xi^a(x)\wedge\delta\xi^b(x')
&(69)\cr}
$$
\noindent where
$$
\omega_{ab}(x,x')=-{ik\over 2\pi}\left(\matrix{
0&\delta(x-x')\delta^{ab}\cr
-\delta(x-x')\delta^{ab}&0\cr}\right)\eqno(70)
$$
\noindent and $\delta\xi^a=(\delta a_{\bar z}^a,\delta a_z^a)$.
$\omega$ defined by (69,70) is called the
symplectic structure and is of course the analogue of
$dp_idx^i$ of particle mechanics. The inverse of $\omega_{ab}$ gives
the Poisson brackets, the commutators being $i$ times the Poisson
brackets.
For our case we get
$$[\xi^a(x),\xi^b(x')]=i(\omega^{-1})^{ab}(x,x')$$
\noindent or
$$[a^a_{\bar z}(x),a_z^b(x)]={2\pi\over k}\delta^{ab}\delta^{(2)}(x-x').
\eqno(71)
$$
\noindent One does not have to go through the symplectic structure
to arrive at (71). One could simply use the fact that, from
(67) the canonical momenta are $\pi=a_{\bar z}$, ${\bar \pi}=0$.
This is thus a constrained system in the Dirac sense and using the
theory of constraints one can derive (71). The procedure of using
$\omega_{ab}(x,x')$ is quicker.

In the expression (68) for $\theta$, $a_{\bar z}$ is independent
of $a_z$. We can however express $a_{\bar z}$ as a function of
$a_z$ via the constraint (65) or equivalently (66) and functionally
integrate $\theta$. In other words, we define $I(a_z)$ by
$$
\delta I={ik\over\pi}\int d^2x~\tr
[a_{\bar z}(a_z)\delta a_z].\eqno(72)
$$
\noindent The solution for $I$ is given by
$$
I=ik\sum{(-1)^n\over n}\int {d^2z_1\over\pi}\cdots
{d^2z_n\over\pi}~
{\tr(a_z(z_1,{\bar z}_1)\ldots a_z(z_n,{\bar z}_n))\over
{\bar z}_{12}{\bar z}_{23}\ldots{\bar z}_{n-1n}{\bar z}_{n1}}.
\eqno(73)
$$

The quantity $I$ has a rather simple interpretation. For
one-dimensional point-particle mechanics, $\theta$, as
we mentioned earlier, is given by $pdx$. $p$ is independent
of $x$ to begin with, but we can express it as a function of
$x$ via a constraint such as of fixed energy, e.g.
${p^2\over2m}+V(x)=E$. Integral of
$\theta=pdx$ then gives Hamilton's principal function or the
eikonal, familiar as the exponent for the WKB wave functions
of one-dimensional quantum mechanics. We have an
analogous situation with (65) expressing $a_{\bar z}$ as a
function of $a_z$. $I$ is thus an eikonal of the Chern-Simons
theory.

$I$ is in fact the Wess-Zumino-Witten action $^{18}$. We can write the
gauge potential $a_z$ as $a_z=-\partial_zUU^{-1}$ where $U$ is
in general not unitary; it is an $SL(N,{\bf C})$ matrix for
gauge group $SU(N)$. Notice that since $\partial_z$ has
an inverse by virtue of $\partial_z{1\over ({\bar z}-{\bar z}')} =\pi
\delta^{(2)}(z - z')$,
such a $U$ can be constructed for any $a_z$, at least as a power
series in $a_z$. $I$ can then be written as $I=-ikS_{WZW}(U)$
where
$$
S_{WZW}(U)={1\over2\pi}\int_Md^2x~\tr(\partial_zU\partial_{\bar z}
U^{-1})-{i\over 12\pi}
\int_{M^3}d^3x~\epsilon^{\mu\nu\alpha}\tr(U^{-1}\partial_\mu U
U^{-1}\partial_\nu U U^{-1}\partial_\alpha U).\eqno(74)
$$
\noindent As usual the second term involves an extension of $U$ into
a three-dimensional space. We take $M^3=M\times [0,1]$ with
$U(z,{\bar z},0)=1$, $U(z,{\bar z},1)=U(z,{\bar z})$. The relationship
to $I$ is easily seen by considering variations of $S_{WZW}$. Under
the variation $U\rightarrow e^\varphi U\simeq(1+\varphi)U$,
we find $\delta a_z=-D_z\varphi=-(\partial_z\varphi+
[a_z,\varphi])$ and
$$
\delta S_{WZW}={1\over\pi}\int d^2x~\tr(\partial_{\bar z}\varphi a_z).
\eqno(75)
$$
\noindent Partially integrating and using $F_{{\bar z}z}=0$
and $\delta a_z=-D_z\varphi$, we find that
$S_{WZW}$ obeys (72) except for a factor $(-ik)$, thus
identifying $I=-ikS_{WZW}$.

Another quantity of geometrical significance for the CS
theory is the K\"{a}hler potential. The phase space of the CS
theory, viz. the function-space of the potentials $a_z$, $a_{\bar z}$
is actually a K\"{a}hler manifold. i.e. it is a complex manifold
with a metric or distance function defined in terms of a potential.
Specifically,
$$\parallel \delta a \parallel^2={k\over 2\pi}\int_M d^2x~
\delta a_{\bar z}^a \delta a^a_z  =
\int d^2x d^2x' ~\left[~{\delta\over\delta a^a_{\bar z}(x)}
{\delta\over\delta a_z^b(x')}K\right]~
\delta a_{\bar z}^a(x)\delta a_z^b(x').
\eqno(76a)$$
\noindent The potential functional $K$ for the metric is the
K\"{a}hler form. Alternatively we may define $K$ by
$$
\omega = -i\int d^2x d^2x' ~
\left[~{\delta\over\delta a^a_{\bar z}(x)}{\delta\over\delta a_z^b(x')}K
\right]~
\delta a^a_{\bar z}(x) \wedge \delta a^b_z (x').
\eqno(76b)
$$
\noindent We find that
$$K={k\over2\pi}\int d^2x ~a_{\bar z}^a a^a_z  + h(a_{\bar z})
+{\overline {h(a_{\bar z})}}\eqno(77)$$
\noindent where $h$ is an arbitrary function of the argument indicated.
In general, there is nothing to dictate any preferred choice for
$h$; but for the CS theory, because there is the additional
structure of gauge transformations, we can choose $h$ so as to
make $K$ gauge invariant. $h$ is then proportional to the
eikonal $I$ and the gauge invariant K\"ahler potential is

$$K=-{1\over \pi}\left[~k\int d^2x~ \tr (a_{\bar z} a_z)
+i\pi I(a_z) +i\pi {\tilde I}(a_{\bar z})\right].\eqno(78)
$$

Finally notice that the eikonal $I$ of (72) may be regarded
as the expansion in powers of $a_z$ of the logarithm of
the functional determinant of $D_z=\partial_z+a_z$; i.e. $I=(-ik)\log\det
D_z=-ik~\Tr\log D_z$. The K\"ahler potential $K$ of (78)
is then given by $-k\Tr\log(D_zD_{\bar z})$.
The expansion of this expression in powers of the potential
obviously gives $I$ and ${\tilde I}$. The extra term
$\int{1\over \pi}\tr(a_{\bar z}a_z)$ is precisely the local counterterm
needed to give a gauge invariantly regulated meaning to
$\Tr\log(D_zD_{\bar z})$ $~^{19}$.
\vskip .2in
\noindent
{\bf 5. The Action for the Hot Quark-Gluon Plasma}
\vskip .2in
We now return to equations (61,62) for the quark-gluon plasma.
We can rewrite (61) as
$$
\delta W=4\int d^4x ~\tr(a_{\bar z}\delta a_z)-\int d^4x~ a_z^a a_z^a .
\eqno(79)
$$
\noindent Comparing with (72), we see that, since $a_{\bar z}$, $a_z$ obey
the constraint (62), the solution is related to the eikonal $I$.
The difference here is that $a_z=A_{+}$ and $a_{\bar z}$ depend on all four
coordinates $x_\mu$, not just $z$, ${\bar z}$. However, there are no
derivatives with respect to the transverse coordinates $x^T$ in (62)
and hence the solution for $a_{\bar z}$ in terms of $a_z$ is the
same as in (66), with $a_z$ depending on $x^T$ in addition to
$z$, ${\bar z}$. The argument of all $a_z$ factors is the
same for $x^T$, i.e.
$$
a_{\bar z} = \sum (-1)^{n-1}\int {d^2z_1\over\pi}\cdots{d^2z_n\over\pi}
{}~{a_z(z_1,{\bar z}_1,x^T)a_z(z_2,{\bar z}_2,x^T)\ldots a_z(z_n,{\bar
z}_n,x^T)
\over ({\bar z}-{\bar z}_1){\bar z}_{12}\ldots {\bar z}_{n-1n}
({\bar z}_n-{\bar z})}.
\eqno(80)
$$
\noindent For the eikonal we get the same expression as (73) but with
integration over the transverse coordinates, i.e.
$$
I=ik\sum {(-1)^n\over n} \int d^2x^T~
{d^2z_1\over\pi}\ldots {d^2z_n\over\pi}~
{\tr (a_z(x_1)\ldots a_z(x_n))\over {\bar z}_{12}{\bar z}_{23}\ldots
{\bar z}_{n1}}.\eqno(81)
$$
\noindent Since it is not relevant to the present discussion, we have,
for the moment, set $k=1$. We then find the solution to (79) as
$$
W=-4\pi i I(a_z)-\int a_z^a a_z^a d^4x.\eqno(82)
$$
\noindent From (55), $\Gamma$ is given by
$$
\Gamma=(N+ {\textstyle {1\over 2}}N_F){T^2\over 12\pi}\left[\int d^4x~ 2\pi
A_0^aA_0^a -\int d^4x d\Omega ~A_{+}^a A_{+}^a -4\pi i I(A_{+})\right].
\eqno(83)
$$
\noindent We can now use identity (45) in reverse to write
$$\eqalignno{\Gamma&=-(N+{\textstyle {1\over 2}}N_F) {T^2\over 6\pi}
\int d\Omega
\left[\int d^4x~\tr(A_{+}A_{-}) +i\pi I(A_{+}) +i\pi \tilde I(A_{-})
\right]&(84a)\cr
&=(N+{\textstyle {1\over 2}}N_F){T^2\over 6}\int d\Omega
{}~K(A_{+}, A_{-})&(84b)\cr}
$$
\noindent where $K$ is given by (78) with the additional integration over
coordinates ${\vec x}^T$ transverse to ${\vec Q}$. If we write $A_{+}=
-\partial_z U U^{-1}$ and $A_{-}={U^{\dagger}}^{-1}\partial_{\bar z} U$,
we can write (84) in terms of $S_{WZW}$ as
$$
\Gamma = -(N+{\textstyle {1\over 2}}N_F){T^2\over 6}S_{WZW}(U^{\dagger}U).
\eqno(85)
$$
\noindent (Again, a suitable additional integration over the transverse
coordinates is understood.)

 From equation (55) or (59) it may seem that our solution
is ambiguous up to the addition of a purely gauge invariant
term. But for hard thermal loops, the additional structure that $W$
depends only on $a_z={1\over 2}A\cdot Q$ tells us that the gauge
invariant piece must obey $D_z{\delta W\over \delta a_z}=0$.
Since $\partial_z$ is invertible, at least
perturbatively, there is no nontrivial solution to
this equation. Thus (83) or (84) is the unique solution and $\Gamma$ so
defined must indeed be the generator of hard thermal loops.

In the last section, we noted that $K(A_+ , A_- )$ can be considered as
$\Tr {\rm log}(D_z D_{\bar z})$. Since $D_z$ and $D_{\bar z}$ are the
chiral Dirac operators in two dimensions, $\Tr {\rm log}(D_z D_{\bar
z})$ is clearly the photon mass term of the Schwinger model, for Abelian
gauge fields. More generally, for non-Abelian fields as well, we can
consider $\Tr {\rm log}(D_z D_{\bar z})$ as a gauge invariant mass term.
It is perhaps fitting that the gauge invariant Debye screening mass term
in four-dimensional QCD is given by suitable integrations of such a
two-dimensional mass term (with, of course, the additional
$x^T$-dependence).

We close this section with two remarks on $\Gamma$. The
CS theory (63) violates parity. We see that this parity violation
disappears, as indeed it should, by integration over
the orientations of ${\vec Q}$, for the QCD case.
Alternatively, expression (84a)
is manifestly parity-symmetric with $A_{+}\leftrightarrow A_{-}$,
$Q\leftrightarrow Q'$ under parity. Secondly, instead of setting $k=1$
and then having a prefactor $(N+{1\over 2}N_F){T^2\over6}$ in
equation (84), we could simply choose $k=(N+{1\over 2}N_F){T^2\over6}$.
For WZW actions $S_{WZW}(U)$ for which $U$ has a non-Abelian
unitary part, any action we use must be an integer times $S_{WZW}$.
This has to do with the fact that there are homotopically nontrivial
maps from the three-sphere $S^3$ into the space of matrices $U$, characterized
by the fact that the third homotopy group $\Pi_3$ of the space of
matrices $U$ is the set of integers $^{18}$.
In our case, $\Gamma$ involves $S_{WZW}(U^{\dagger}U)$ or
$S_{WZW}(H)$ where $H$ is hermitian. The space of hermitian
matrices has trivial $\Pi_3$ and so, as expected, there is no
argument for quantization of the coefficient, which
is $(N+{1\over 2}N_F){T^2\over6}$ for us.

We have given the coefficient of $K$ in (84) for a plasma
with zero chemical potential. For the more general case, we can
write $\Gamma$ as
$$
\Gamma=\int {1\over 2q}{d^3 q\over (2\pi)^3}
[Nn_g+\sum^{N_F}_{i=1}({n_i+{\bar n}_i\over2})]
{}~\left[\int d^4x~\tr(A_{+}A_{-}) +i\pi I(A_{+}) + i\pi{\tilde I}(A_{-})
\right].
\eqno(86)$$
\vskip .2in
\noindent
{\bf 6. Plasma Waves and Debye Screening}
\vskip .2in
We have obtained the effective action $\Gamma$ which is the
generator of hard thermal loops. As is standard in other contexts of
resummations of perturbation theory, we can develop the resummed
perturbative expansion of thermal QCD as follows. We introduce a splitting
of the Yang-Mills action as
$$\eqalignno{&S=S_0-c\Gamma\cr
	     &S_0=\int ~{\textstyle-{1\over 4}}F^2~+\Gamma&(87)\cr}
$$
\noindent We define propagators and vertices and start off the
perturbative expansion using $S_0$. $c\Gamma$ will be
treated
as a `counterterm', nominally one order higher in the thermal loops
than $S_0$. Eventually of course $c$ is taken to be 1, so that
we are only achieving a rearrangement of terms in the
perturbative expansion. (As usual we must have gauge fixing and ghost terms.
We have also not displayed the quark terms; for the quark terms, see
ref.20.) Using this procedure one can
calculate quantities which require resummations such as the gluon
decay rate in the plasma $^4$.

$\Gamma$ also gives us an effective action for low energy gluon fields
$$
S_{eff}=\int ~{\textstyle-{1\over4}}F^2~+\Gamma.
\eqno(88)
$$
\noindent (Although $S_{eff}$ looks like $S_0$ in (87), the interpretation
is different. $S=S_0-c\Gamma$ in (87) generates the thermal
perturbation theory. Momenta for fields in $S_0$, for example, can be
very high. Calculations starting with (87) lead to a low energy effective
action (88); the latter is, of course, useful only for small momenta.)

Long wavelength and low frequency plasma waves are the classical solutions
of the effective theory (88). It also includes effects such as the screening
of Coulomb fields. These features can be seen by examining the Abelian
case or electrodynamics.
In this case, the terms in $\Gamma$ which are cubic or higher
order in $A_\mu$ are zero and in terms of the Fourier components of $A_\mu$,
we can write
$$
S_{eff}={\textstyle{1\over2}}\int A_\mu(-k)M^{\mu\nu}(k) A_\nu(k)
{d^4k\over(2\pi)^4}
\eqno(89a)
$$
\noindent where
$$
M^{\mu\nu}=(-k^2g^{\mu\nu}+k^\mu k^\nu)+
{N_Fe^2T^2\over 12\pi}
\left[4\pi\delta^{\mu 0}\delta^{\nu 0}
-\int d\Omega ~{k_0\over k\cdot Q}Q^\mu Q^\nu\right].
\eqno(89b)
$$
$k^\mu M_{\mu\nu}=0$ in accordance with the requirement of gauge invariance.
We have restored the coupling constant $e$ at this stage.
We can split $A_\mu$ into a gauge dependent part and gauge
invariant components as
$$A_\mu =k_\mu \Lambda (k) +\alpha_\mu +\beta_\mu\eqno(90)
$$
\noindent where $\Lambda$ shifts under gauge transformations and
$\alpha_\mu$, $\beta_\mu$ are gauge invariant. We take
$$
\alpha_0=({{\vec k}^2\over {\vec k}^2-k_0^2})\phi ~~~~
\alpha_i={k_0\over\sqrt{{\vec k}^2}}e^{(3)}_i~({{\vec k}^2\over
{\vec k}^2-k_0^2})\phi
$$
$$\beta_0=0~~~~
\beta_i=e_i^{(\lambda)}a_\lambda, ~~~~\lambda=1,2.\eqno(91)
$$
(Here we are considering fields off-shell and so $k^0 \neq \vert \vec k
\vert$.) The $e_i$'s form a triad of spatial unit vectors which
may be taken as
$$e_i^{(3)}={k_i\over\sqrt{{\vec k}^2}},~~~~i=1,2,3,$$
$$e^{(1)}=(\epsilon_{ij}{k_j\over\sqrt{k_T^2}},0),~
{}~~~e^{(2)}=({k_3k_i\over\sqrt{k_T^2{\vec k}^2}},-\sqrt{k_T^2\over
{\vec k}^2}), ~~~~i=1,2, \eqno(92)
$$
\noindent where $k_T^2=k_1^2+k_2^2$. Notice that $k_ie_i^{(\lambda)}=0$,
$\lambda=1,2$. $\phi$ and $a_\lambda$ are the gauge invariant degrees
of freedom in (90). When the mode
decomposition (90) is used in (89) we get
$$
S_{eff}={\textstyle{1\over2}}\int {d^4k\over (2\pi)^4}
{}~\left[~\beta_i (-k)({{\vec k}^2\delta_{ij}-k_ik_j\over{\vec k}^2})
M^T(k)\beta_j(k)+\phi(-k)M^L(k)\phi(k)\right]
+\int\phi J^0+\beta_i J^i\eqno(93)
$$
\noindent where
$$M^T(k)=k_0^2-{\vec k}^2-{N_Fe^2 T^2\over6}\left[~{k_0^2\over{\vec k}^2}
+(1-{k_0^2\over{\vec k}^2}){k_0\over 2|{\vec k}|}L~\right]$$
$$M^L(k)={\vec k}^2+{N_F e^2T^2\over3}(1-{k_0\over2|{\vec k}|}L)\eqno(94)
$$
$$L=\log({k_0+k\over k_0-k}).\eqno(95)$$
\noindent We have also included an interaction term with a source
$J_\mu$ in (93); i.e. we include $\int A_\mu J^\mu$ and simplify it
using (90). From (93) we see that the interaction between
charges in the plasma is governed by $(M^L)^{-1}$ which shows the
Debye screening with a Debye mass $m_D=\sqrt{N_Fe^2 T^2\over 3}$.
The action (93) can also give free wavelike solutions.
The dispersion rules for these plasma waves would
be $M^T=0$ for the transverse waves and $M^L=0$ for the
longitudinal waves $^{21}$.

Non-Abelian plasma waves can be defined in a similar way as propagating
solutions to the equations of motion given by (88). If one
truncates $\Gamma$ to the term quadratic in $A_\mu$, these
are essentially the same as the Abelian plasma waves with
$m_D^2=(N+{1\over 2}N_F){g^2T^2\over3}$ and $(N+{1\over 2}N_F){g^2T^2\over6}$
in $M^T(k)$ rather than ${N_Fe^2 T^2\over6}$. However such
a truncation of $\Gamma$ loses the full gauge invariance. In the
approximation of small $gA_\mu$, i.e. oscillations of very small
amplitudes, this lack of gauge invariance may not be very serious.
However, in general, one
must really seek propagating solutions to the
equations of motion keeping all of $\Gamma$. These will
be the genuine
plasma oscillations in the non-Abelian case.
We do not have any such solutions as yet, but some of them might
coincide with those found by Kajantie and Montonen $^{22}$.

\vskip .3in
\noindent
{\bf References}
\vskip .2in
\item
{1.} R.Efraty and V.P.Nair, {\it Phys.Rev.Lett} {\bf 68}, 2891 (1992).
\vskip .1in
\item
{2.}V.P.Silin, {\it Sov.J.Phys.JETP} {\bf 11}, 1136 (1960);
V.V.Klimov, {\it Sov.J.Nucl.Phys.} {\bf 33}, 934 (1981);{\it Sov.Phys.JETP}
{\bf 55}, 199 (1982);
H.A.Weldon, {\it Phys.Rev.D} {\bf 26}, 1394 (1982).
\vskip .1in
\item
{3.} R.Pisarski, {\it Physica} {\bf A158}, 246 (1989);
{\it Phys.Rev.Lett.} {\bf
63}, 1129 (1989).
\vskip .1in
\item
{4.} E.Braaten and R.Pisarski, {\it Phys.Rev.} {\bf D42},
2156 (1990); {\it Nucl.Phys.} {\bf B337}, 569 (1990); {\it ibid} {\bf B339},
310 (1990); {\it Phys.Rev.} {\bf D 45},1827 (1992).
\vskip .1in
\item
{5.} R.Jackiw and S.Templeton, {\it Phys.Rev.} {\bf D23}, 2291 (1981);
J.Schonfeld, {\it Nucl.Phys.} {\bf B185}, 157 (1981); S.Deser, R.Jackiw
and
S.Templeton, {\it Phys.Rev.Lett.} {\bf 48}, 975 (1982); {\it Ann.Phys.}
{\bf
140}, 372 (1982).
\vskip .1in
\item
{6.} See for example, S.Forte, {\it Rev.Mod.Phys.} {\bf 64}, 193 (1992);
R.Jackiw, in {\it Physics, Geometry and Topology}, Proceedings of the
NATO ASI, Banff 1989, H.C.Lee (ed.), Plenum Press (1990).
\vskip .1in
\item
{7.} E.Witten, {\it Comm.Math.Phys.} {\bf 121}, 351 (1989).
\vskip .1in
\item
{8.} M.Bos and V.P.Nair, {\it Phys.Lett.} {\bf B223}, 61 (1989); {\it
Int.J.Mod.Phys.} {\bf A5}, 959 (1990); S.Elitzur, G.Moore, A.Schwimmer
and
N.Seiberg, {\it Nucl.Phys.} {\bf B326}, 108 (1989); J.M.F.Labastida and
A.V.Ramallo, {\it Phys.Lett.} {\bf B227}, 92 (1989); H.Murayama, {\it
Z.Phys.}
{\bf C48}, 79 (1990); A.P.Polychronakos, {\it Ann.Phys.} {\bf 203}, 231
(1990);
T.R.Ramadas, I.M.Singer and J.Weitsman, {\it Comm.Math.Phys.} {\bf 126},
409 (1989); A.P.Balachandran, M.Bourdeau and S.Jo {\it Mod.Phys.Lett.}
{\bf A 4}, 1923 (1989); {\it Int.J.Mod.Phys.} {\bf A 5}, 2423 (1990);
{\bf 5}, 3461 (1990); A.P.Balachandran, G.Bimonte, K.S.Gupta and A. Stern,
{\it Int.J.Mod.Phys.} {\bf A7}, 4655 (1992).
\vskip .1in
\item
{9.} R.Jackiw and E.Weinberg, {\it Phys.Rev.Lett.} {\bf 64}, 2234 (1990);
R.Jackiw, K.Lee and E.Weinberg, {\it Phys.Rev.} {\bf D42}, 3488 (1990);
J.Hong,
Y.Kim and P.Y.Pac, {\it Phys.Rev.Lett.} {\bf 64}, 2230 (1990);
S.Paul and A.Khare, {\it Phys.Lett} {\bf 174B}, 420 (1986); (E) {\bf 182B},
414 (1986); L.Jacobs, A.Khare, C.Kumar and S.Paul, {\it Int.J.Mod.Phys.}
{\bf A6}, 3441 (1991).
\vskip .1in
\item
{10.} V.P.Nair and J.Schiff, {\it Phys.Lett.} {\bf B246}, 423 (1990);
{\it Nucl.Phys.} {\bf B371}, 329 (1992).
\vskip .1in
\item
{11.} J.Frenkel and J.C.Taylor, {\it Nucl.Phys.} {\bf B334}, 199 (1990).
\vskip .1in
\item
{12.} J.C.Taylor and S.M.H.Wong, {\it Nucl.Phys.} {\bf B346}, 115 (1990).
\vskip .1in
\item
{13.} N.P.Landsman and Ch.G.van Weert, {\it Phys.Rep.} {\bf 145}, 141 (1987).
\vskip .1in
\item
{14.} G.Barton, {\it Ann.Phys.(N.Y.)} {\bf 200}, 271 (1990).
\vskip .1in
\item
{15.}
R.Kobes, G.Kunstatter and A.Rebhan, {\it Nucl.Phys.} {\bf B355}, 1 (1991).
\vskip .1in
\item
{16.}D.Gonzales and A.N.Redlich, {\it Ann.Phys.(N.Y.)}
{\bf 169}, 104 (1986);
B.M.Zupnik, {\it Phys.Lett.} {\bf B183}, 175 (1987); G.V.Dunne, R.Jackiw
and C.A.Trugenberger, Ann.Phys.(N.Y.) {\bf 149}, 197 (1989).
\vskip .1in
\item
{17.} V.Guillemin and S.Sternberg {\it Symplectic Techniques in Physics},
Cambridge University Press (1990); J.Schwinger, Phys.Rev. {\bf 82}, 914
(1951);
C.Crnkovic and E.Witten, in {\it Three Hundred Years of
Gravitation},
S.W.Hawking and W.Israel (eds.), Cambridge University Press (1987);
G.J.Zuckerman, in {\it Mathematical Aspects of String Theory}, S.T.Yau (ed.),
World Scientific (1987); L.Faddeev and R.Jackiw, {\it Phys.Rev.Lett.}
{\bf 60}, 1692 (1988).
\vskip .1in
\item
{18.} E.Witten, {\it Comm.Math.Phys.} {\bf 92}, 455 (1984).
\vskip .1in
\item
{19.} A.M.Polyakov and P.B.Wiegmann, Phys.Lett. {\bf 141B}, 223 (1984);
D.Karabali, Q-H Park, H.J.Schnitzer and Z.Yang, {\it Phys.Lett.}
{\bf 216B}, 307 (1989); D.Karabali and H.J.Schnitzer, {\it Nucl.Phys.}
{\bf B329}, 649 (1990); K.Gawedzki and A.Kupianen, {\it Phys.Lett.}
{\bf 215B}, 119 (1988); {\it Nucl.Phys.}, {\bf B320}, 649 (1989).
\vskip .1in
\item
{20.} E.Braaten, in {\it Hot Summer Daze}, A.Gocksch and R.Pisarski (eds.),
World Scientific (1992).
\vskip .1in
\item
{21.} R.Pisarski, {\it Physica} {\bf A158}, 146 (1989); T.Appelquist and
R.Pisarski, {\it Phys.Rev}, {\bf D23} (1981).
\vskip .1in
\item
{22.} K.Kajantie and C.Montonen, {\it Physica Scripta} {\bf 22}, 555 (1980).

\end